\documentclass[runningheads]{llncs}
\usepackage[T1]{fontenc}
\usepackage{graphicx}
\usepackage{booktabs}
\usepackage{siunitx}
\usepackage{amsmath}
\usepackage{amssymb}
\usepackage{multirow}
\usepackage{hyperref}

\usepackage{xcolor}
\usepackage{sidecap}
\usepackage[misc]{ifsym}

\hypersetup{
    colorlinks,
    linkcolor={blue!50!black},
    citecolor={blue!50!black},
    urlcolor={blue!50!black}
}

\begin{document}
\title{Robustness Stress Testing in Medical Image Classification}
%

\author{Mobarakol Islam\inst{1,2}$^{\textrm{(\Letter)}}$, Zeju Li\inst{1,3}, and Ben Glocker\inst{1}}
\institute{BioMedIA Group, Department of Computing, Imperial College London, UK \and  Wellcome/EPSRC Centre for Interventional and Surgical Sciences(WEISS) and Dept. of Medical Physics and Biomedical Engineering, University College London, UK \and FMRIB Centre, Nuffield Dept. of Clinical Neurosciences, University of Oxford, UK\\ 
\email{mobarakol.islam@ucl.ac.uk} }

\maketitle              
\begin{abstract}
Deep neural networks have shown impressive performance for image-based disease detection. Performance is commonly evaluated through clinical validation on independent test sets to demonstrate clinically acceptable accuracy. Reporting good performance metrics on test sets, however, is not always a sufficient indication of the generalizability and robustness of an algorithm. In particular, when the test data is drawn from the same distribution as the training data, the iid test set performance can be an unreliable estimate of the accuracy on new data. In this paper, we employ stress testing to assess model robustness and subgroup performance disparities in disease detection models. We design progressive stress testing using five different bidirectional and unidirectional image perturbations with six different severity levels. As a use case, we apply stress tests to measure the robustness of disease detection models for chest X-ray and skin lesion images, and demonstrate the importance of studying class and domain-specific model behaviour. Our experiments indicate that some models may yield more robust and equitable performance than others. We also find that pretraining characteristics play an important role in downstream robustness. We conclude that progressive stress testing is a viable and important tool and should become standard practice in the clinical validation of image-based disease detection models\footnote{ Source code: \url{https://github.com/mobarakol/Robustness_Stress_Testing}}.

\end{abstract}
\section{Introduction}
Despite expert-level performance of artificial intelligence (AI) systems on some image-based disease detection tasks \cite{gulshan2016development,de2018clinically,esteva2017dermatologist}, there remain concerns regarding the generalizability and robustness of such systems when deployed in clinical practice \cite{finlayson2020clinician}. The systems may need to process new data with different characteristics compared to the development data. The performance on such new data may be different from the one observed previously during clinical validation, in particular, when the previous test data was drawn from the same distribution as the training data. This commonly occurs in controlled experimental settings where the final model performance is established on an independent and identically distributed (iid) test set. But even when external datasets are being used as part of the validation, the observed performance in terms of true and false positive rates (TPR/FPR), or threshold-agnostic metrics such as area under the receiver operating characteristic curve (AUC), may be different after deployment when the data characteristics change. Such performance drift of AI models can occur due to changes in the image acquisition, upgrades to scanner hardware, or other distribution shifts, for example, in the patient population~\cite{castro2020causality}. The performance drift between the development and deployment stage with robustness concern has been associated with (i) spurious correlations between the training data and the output target labels~\cite{saab2022reducing}, and (ii) underspecification~\cite{d2020underspecification} where deep neural networks are notoriously underspecified regarding the `to be learned mechanism' mapping inputs to outputs. While two models may perform similarly well on the majority of patients, there can be significant differences in the performance across subgroups \cite{oakden2020hidden} or underrepresented populations \cite{seyyed2020chexclusion,seyyed2021underdiagnosis}. In order to identify and differentiate between models that otherwise have similar iid test set performance, stress testing was proposed as a valuable tool for analysing model robustness \cite{d2020underspecification,eche2021toward}. We argue that stress testing needs to extend to subgroup performance analysis if the goal is to deliver equitable solutions that work across the entire patient population.

Computational stress testing is a well known tool in AI to assess model performance by deliberately modifying the input data to simulate various real-world conditions. Most recently, there are attempts to design stress testing for the assessment of deep neural networks in medical applications~\cite{eche2021toward,young2021stress,yao2021hierarchical,araujo2021stress}. Inspired by~\cite{d2020underspecification}, the study~\cite{eche2021toward} discusses a framework for identifying model underspecification using stress testing by stratifying test sets according to input perturbations. Stress testing can be done with a variety of clinically meaningful transformations to assess model performance beyond the commonly reported metrics~\cite{young2021stress}. The study presented in \cite{young2021stress} finds that a model with acceptable performance on conventional metrics may not necessarily pass the stress test. Adversarial attacks are also found to be utilized in stress testing to reveal the vulnerability of medical images compared to natural images~\cite{yao2021hierarchical}. However, most of these works investigate overall, aggregated performance changes by stratifying test data according to the input perturbation only, ignoring potential performance disparities across subgroups and input domains.

In this work, we highlight the importance of extending stress testing to include subgroup performance analysis. We use two real-world applications of chest X-ray disease detection and skin lesion classification for which we design meaningful stress tests using uni- and bi-directional image perturbation with multiple levels of perturbation severity. We use systematic changes to the image appearance in order to modify the iid test set as illustrated in Fig~\ref{fig:perturbation_techs}(a). This allows us to investigate model robustness and analyse subgroup disparities of disease detection models. We employ {four} different neural network architectures, comparing state-of-the-art convolutional neural networks (DenseNet-121 \cite{huang2017densely} {and ResNet-34~\cite{he2016deep}}) with the emerging vision transformers (ViT \cite{dosovitskiy2020image} {and Swin-Transformer~\cite{liu2021swin}}). We measure the classification accuracy in terms of TPR and FPR for a specific operating point, and the general ability of the models to identify disease via the threshold-agnostic AUC metric. Our key findings are (i) Stress testing is capable of identifying performance differences beyond iid test set accuracy; (ii) Transformer-based models may potentially be more robust compared to traditional CNNs; (iii) Subgroup performance should be analysed using class-specific stress testing to check for disparities; (iv) Stress testing reveals that changes in performance across subgroups can relate to jointly shifting TPR and FPR while the AUC remains similar across groups. The last observation is of particular importance as it raises questions about the clinical utility of these models. A model that shows a shift in TPR/FPR but performs equally well in terms of AUC on different subgroups requires careful analysis of the underlying causes of the performance shift, which could be related to various types of dataset and model bias \cite{bernhardt2022potential}.

\section{Stress testing via image perturbations}

\begin{figure}[!h]
\centering
\includegraphics[width=0.85\textwidth]{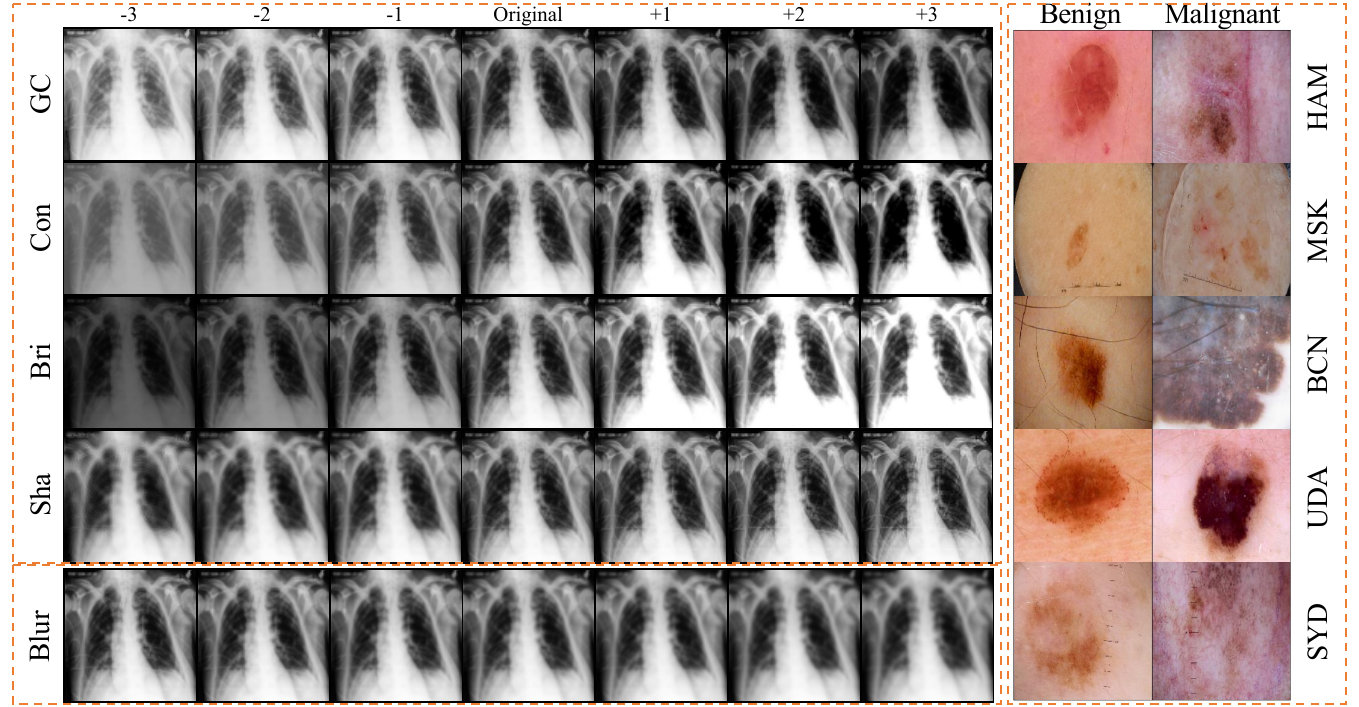}
\caption{Stress testing via image perturbations: (left) Effects of different perturbations shown for an example chest X-ray image; (right) Benign and malignant images from five different sites with domain shift in ISIC skin lesion dataset.}
\centering
\label{fig:perturbation_techs}
\end{figure}

In stress testing, model performance is assessed by applying meaningful perturbations to the input images. In this way, a model can be tested systematically to measure the robustness under simulated yet realistic conditions. Perturbation techniques such as blurring, changes to the brightness and contrast, or pixelation have been previously employed. In addition to appearance changes, spatial transformations such as horizontal flipping and rotations have been used~\cite{young2021stress}. Several studies have employed extensive perturbations to determine model robustness~\cite{hendrycks2019benchmarking,taori2020measuring,wiles2021fine}. In this work, we design stress tests using four bidirectional and one unidirectional image perturbations with six different severity levels. The perturbations are applied during test time. We use systematic changes via gamma correction (GC), contrast (Con), brightness (Bri), sharpness (Sha), and Gaussian blur (Blur) in order to modify the iid test set. Fig~\ref{fig:perturbation_techs}(a)) visualizes the effect of the perturbations on an example input image with different levels of severity. Positive and negative adjustments can be applied for contrast, brightness, gamma correction, and sharpness. We adopt the implementation of these transformations from the torchvision functional library which facilitates straightforward reproducibility of our experiments\footnote{\url{https://pytorch.org/vision/stable/transforms.html}}. The choice of suitable stress tests is data and task-specific. We consider perturbations that are meaningful for X-ray and skin lesion imaging {and directly relate to key aspects of any image acquisition. Other perturbations may be considered for providing a more complete picture of the model performance. For a specific application and clinical use, the stress tests should be designed according to the characteristics of the application.}

\textbf{Use case 1 (Chest X-ray disease detection):} We utilise the publicly available CheXpert dataset~\cite{irvin2019chexpert} as the method development domain and the MIMIC-CXR dataset~\cite{johnson2019mimic} as the external validation domain. In CheXpert, there are 127,118 chest X-ray scans with 14 different disease multi-label annotations and split into training (76,205), validation (12,673), and testing (38,240) following the setup in~\cite{gichoya2022ai}. The dataset also contains demographic information including racial identity (78\% White, 15 \% Asian, 7\% Black) and biological sex (41\% female and 59\% of male). In MIMIC-CXR~\cite{johnson2019mimic}, there are 183,207 chest X-ray scans with 14 disease annotation as CheXpert and 30\% of the scans are used for external testing in this work. The dataset contains racial identity and biological sex information of 77\% White, 4\% Asian, 19\% Black, 52\% Female and 48\% Male.

\textbf{Use case 2 (Skin lesion classification):} We use the publicly available ISIC~\cite{cassidy2022analysis} skin lesion detection dataset with 52,710 images consisting complete meta-information from 5 different sites (HAM, MSK, BCN, UDA, and SYD) as shown in Fig~\ref{fig:perturbation_techs}(b) and split into 60\% / 40\% for training and testing. We defined five subgroups of male (48\%), female (52\%), young (24\%, age $<$ 39), mid-aged (47\%, 39 $<$ age $>$ 59) and old (29\%, age $>$ 59).

\begin{table}[]
\centering
\caption{Performance of the disease detection
in chest X-ray on iid test set (CheXpert), external test set (MIMIC-CXR) and skin lesion on multi-site test set(ISIC) for DenseNet and ViT.}
\label{tab:chest_xray}
\scalebox{0.75}{
\begin{tabular}{c|c|c|cccccc|cccccc|cccc}
\hline
 &  &  & \multicolumn{6}{c|}{CheXpert} & \multicolumn{6}{c|}{MIMIC-CXR} & \multicolumn{4}{c}{\multirow{2}{*}{ISIC Skin Lesion}} \\ \cline{1-15}
 &  &  & \multicolumn{3}{c|}{No Finding} & \multicolumn{3}{c|}{Pleural Effusion} & \multicolumn{3}{c|}{No Finding} & \multicolumn{3}{c|}{Pleural Effusion} & \multicolumn{4}{c}{} \\ \hline
 &  & Subgrp & \multicolumn{1}{c|}{\begin{tabular}[c]{@{}c@{}}w \\ pretr.\end{tabular}} & \multicolumn{1}{c|}{\begin{tabular}[c]{@{}c@{}}w/o \\ pretr.\end{tabular}} & \multicolumn{1}{c|}{\begin{tabular}[c]{@{}c@{}}w\\ aug.\end{tabular}} & \multicolumn{1}{c|}{\begin{tabular}[c]{@{}c@{}}w \\ pretr.\end{tabular}} & \multicolumn{1}{c|}{\begin{tabular}[c]{@{}c@{}}w/o \\ pretr.\end{tabular}} & \begin{tabular}[c]{@{}c@{}}w\\ aug.\end{tabular} & \multicolumn{1}{c|}{\begin{tabular}[c]{@{}c@{}}w \\ pretr.\end{tabular}} & \multicolumn{1}{c|}{\begin{tabular}[c]{@{}c@{}}w/o \\ pretr.\end{tabular}} & \multicolumn{1}{c|}{\begin{tabular}[c]{@{}c@{}}w\\ pretr.\end{tabular}} & \multicolumn{1}{c|}{\begin{tabular}[c]{@{}c@{}}w \\ pretr.\end{tabular}} & \multicolumn{1}{c|}{\begin{tabular}[c]{@{}c@{}}w/o \\ pretr.\end{tabular}} & \begin{tabular}[c]{@{}c@{}}w\\ aug.\end{tabular} & \multicolumn{1}{c|}{Subgrp} & \multicolumn{1}{c|}{\begin{tabular}[c]{@{}c@{}}w \\ pretr.\end{tabular}} & \multicolumn{1}{c|}{\begin{tabular}[c]{@{}c@{}}w/o \\ pretr.\end{tabular}} & \begin{tabular}[c]{@{}c@{}}w\\ aug.\end{tabular} \\ \hline
\multirow{10}{*}{\rotatebox{90}{AUC}} & \multirow{5}{*}{\rotatebox{90}{DenseNet}} & White & 0.87 & 0.87 & \multicolumn{1}{c|}{0.87} & 0.86 & 0.86 & 0.86 & 0.82 & 0.82 & \multicolumn{1}{c|}{0.83} & 0.88 & 0.88 & 0.89 & \multicolumn{1}{c|}{Young} & 0.79 & 0.81 & 0.82 \\ 
 &  & Asian & 0.88 & 0.87 & \multicolumn{1}{c|}{0.86} & 0.88 & 0.88 & 0.88 & 0.83 & 0.83 & \multicolumn{1}{c|}{0.83} & 0.88 & 0.89 & 0.89 & \multicolumn{1}{c|}{Mid-aged} & 0.80 & 0.78 & 0.82 \\ 
 &  & Black & 0.88 & 0.89 & \multicolumn{1}{c|}{0.89} & 0.86 & 0.86 & 0.86 & 0.83 & 0.82 & \multicolumn{1}{c|}{0.83} & 0.90 & 0.89 & 0.89 & \multicolumn{1}{c|}{Old} & 0.70 & 0.72 & 0.72 \\ 
 &  & Female & 0.87 & 0.87 & \multicolumn{1}{c|}{0.87} & 0.87 & 0.87 & 0.87 & 0.84 & 0.83 & \multicolumn{1}{c|}{0.84} & 0.89 & 0.90 & 0.90 & \multicolumn{1}{c|}{Female} & 0.79 & 0.78 & 0.81 \\ 
 &  & Male & 0.87 & 0.87 & \multicolumn{1}{c|}{0.87} & 0.86 & 0.86 & 0.87 & 0.81 & 0.82 & \multicolumn{1}{c|}{0.82} & 0.88 & 0.88 & 0.88 & \multicolumn{1}{c|}{Male} & 0.79 & 0.79 & 0.81 \\ \cline{2-19} 
 & \multirow{5}{*}{\rotatebox{90}{ViT}} & White & 0.87 & 0.86 & \multicolumn{1}{c|}{0.88} & 0.87 & 0.86 & 0.87 & 0.82 & 0.81 & \multicolumn{1}{c|}{0.82} & 0.89 & 0.88 & 0.89 & \multicolumn{1}{c|}{Young} & 0.86 & 0.82 & 0.87 \\ 
 &  & Asian & 0.87 & 0.87 & \multicolumn{1}{c|}{0.88} & 0.88 & 0.87 & 0.88 & 0.83 & 0.83 & \multicolumn{1}{c|}{0.84} & 0.89 & 0.88 & 0.89 & \multicolumn{1}{c|}{Mid-aged} & 0.84 & 0.82 & 0.85 \\ 
 &  & Black & 0.89 & 0.88 & \multicolumn{1}{c|}{0.89} & 0.87 & 0.86 & 0.87 & 0.83 & 0.82 & \multicolumn{1}{c|}{0.83} & 0.90 & 0.88 & 0.90 & \multicolumn{1}{c|}{Old} & 0.82 & 0.77 & 0.88 \\ 
 &  & Female & 0.87 & 0.86 & \multicolumn{1}{c|}{0.88} & 0.87 & 0.87 & 0.87 & 0.83 & 0.83 & \multicolumn{1}{c|}{0.84} & 0.90 & 0.89 & 0.90 & \multicolumn{1}{c|}{Female} & 0.86 & 0.82 & 0.87 \\ 
 &  & Male & 0.88 & 0.87 & \multicolumn{1}{c|}{0.88} & 0.87 & 0.86 & 0.87 & 0.81 & 0.81 & \multicolumn{1}{c|}{0.81} & 0.88 & 0.88 & 0.88 & \multicolumn{1}{c|}{Male} & 0.83 & 0.81 & 0.83 \\ \hline
\multirow{10}{*}{\rotatebox{90}{F1}} & \multirow{5}{*}{\rotatebox{90}{DenseNet}} & White & 0.40 & 0.40 & \multicolumn{1}{c|}{0.40} & 0.74 & 0.74 & 0.74 & 0.64 & 0.64 & \multicolumn{1}{c|}{0.65} & 0.68 & 0.69 & 0.69 & \multicolumn{1}{c|}{Young} & 0.43 & 0.44 & 0.43 \\ \cline{3-3}
 &  & Asian & 0.42 & 0.42 & \multicolumn{1}{c|}{0.41} & 0.77 & 0.76 & 0.76 & 0.68 & 0.69 & \multicolumn{1}{c|}{0.69} & 0.69 & 0.70 & 0.70 & \multicolumn{1}{c|}{Mid-aged} & 0.49 & 0.46 & 0.49 \\ 
 &  & Black & 0.46 & 0.47 & \multicolumn{1}{c|}{0.46} & 0.70 & 0.68 & 0.70 & 0.69 & 0.68 & \multicolumn{1}{c|}{0.70} & 0.62 & 0.62 & 0.61 & \multicolumn{1}{c|}{Old} & 0.22 & 0.22 & 0.20 \\ 
 &  & Female & 0.41 & 0.41 & \multicolumn{1}{c|}{0.41} & 0.75 & 0.74 & 0.75 & 0.68 & 0.67 & \multicolumn{1}{c|}{0.68} & 0.68 & 0.69 & 0.68 & \multicolumn{1}{c|}{Female} & 0.38 & 0.37 & 0.38 \\ 
 &  & Male & 0.41 & 0.40 & \multicolumn{1}{c|}{0.40} & 0.74 & 0.74 & 0.74 & 0.63 & 0.64 & \multicolumn{1}{c|}{0.64} & 0.67 & 0.67 & 0.68 & \multicolumn{1}{c|}{Male} & 0.56 & 0.52 & 0.60 \\ \cline{2-19} 
 & \multirow{5}{*}{\rotatebox{90}{ViT}} & White & 0.40 & 0.40 & \multicolumn{1}{c|}{0.41} & 0.75 & 0.74 & 0.75 & 0.63 & 0.63 & \multicolumn{1}{c|}{0.64} & 0.69 & 0.69 & 0.69 & \multicolumn{1}{c|}{Young} & 0.47 & 0.44 & 0.47 \\ 
 &  & Asian & 0.43 & 0.42 & \multicolumn{1}{c|}{0.43} & 0.77 & 0.77 & 0.77 & 0.67 & 0.68 & \multicolumn{1}{c|}{0.69} & 0.71 & 0.69 & 0.70 & \multicolumn{1}{c|}{Mid-aged} & 0.52 & 0.48 & 0.55 \\ 
 &  & Black & 0.46 & 0.48 & \multicolumn{1}{c|}{0.47} & 0.71 & 0.71 & 0.71 & 0.69 & 0.68 & \multicolumn{1}{c|}{0.69} & 0.63 & 0.60 & 0.63 & \multicolumn{1}{c|}{Old} & 0.24 & 0.18 & 0.26 \\ 
 &  & Female & 0.40 & 0.41 & \multicolumn{1}{c|}{0.42} & 0.75 & 0.75 & 0.75 & 0.67 & 0.66 & \multicolumn{1}{c|}{0.68} & 0.69 & 0.68 & 0.69 & \multicolumn{1}{c|}{Female} & 0.43 & 0.41 & 0.43 \\ 
 &  & Male & 0.42 & 0.41 & \multicolumn{1}{c|}{0.42} & 0.75 & 0.74 & 0.75 & 0.63 & 0.62 & \multicolumn{1}{c|}{0.64} & 0.68 & 0.67 & 0.68 & \multicolumn{1}{c|}{Male} & 0.63 & 0.60 & 0.62 \\ \hline
\end{tabular}
}
\end{table}

\section{Experimental setup}

To evaluate the performance of medical image classification under stress testing, we train chest X-ray disease detection and skin lesion classification models employing different types of state-of-the-art deep neural network architectures. We use two popular CNN-based architectures, DenseNet-121~\cite{huang2017densely} and ResNet-34~\cite{he2016deep}, and two recent transformer-based models, the Vision Transformer (ViT)~\cite{dosovitskiy2020image} and Swin-Transformer~\cite{liu2021swin}. ViT is a transformer-like architecture over patches of the image where Swin-transformer introduce hierarchical feature maps and shifted window attention. For training, we employ the Adam optimizer with a learning rate of 0.001 and binary cross-entropy loss for multi-label (14 classes) chest X-ray disease detection and a binary cross-entropy loss for skin lesion classification.

\section{Results and Findings}

\begin{figure*}[!h]
\centering
\includegraphics[width=1\textwidth]{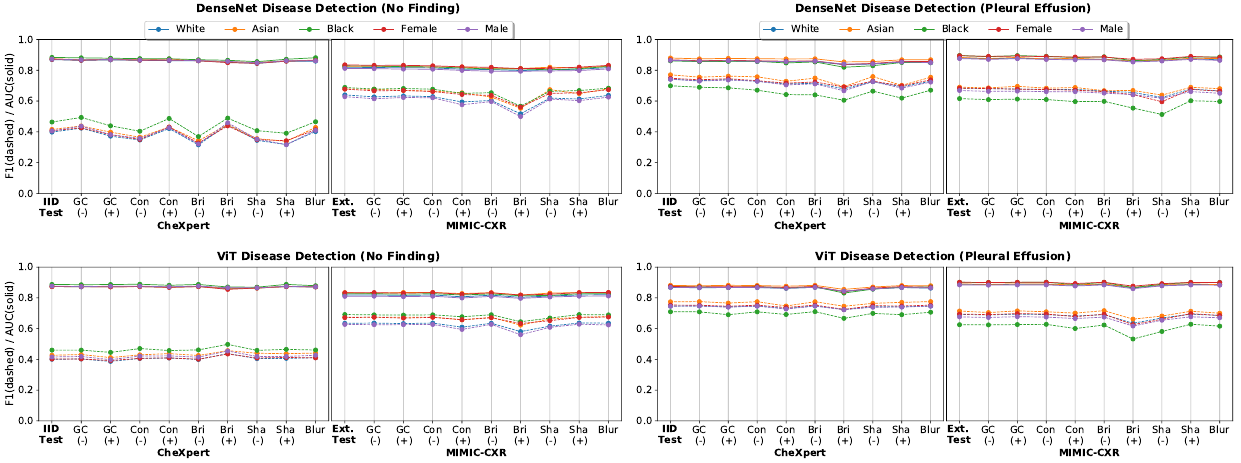}
\caption{F1 and AUC across patient subgroups for a DenseNet (top) and ViT (bottom) disease detection model for a variety of stress tests on the development data (CheXpert) and the external validation data (MIMIC-CXR). Performance for `no-finding' is shown on the left, and `pleural effusion' on the right.}
\centering
\label{fig:robustness_auc}
\end{figure*}

\begin{figure*}[!h]
\centering
\includegraphics[width=1\textwidth]{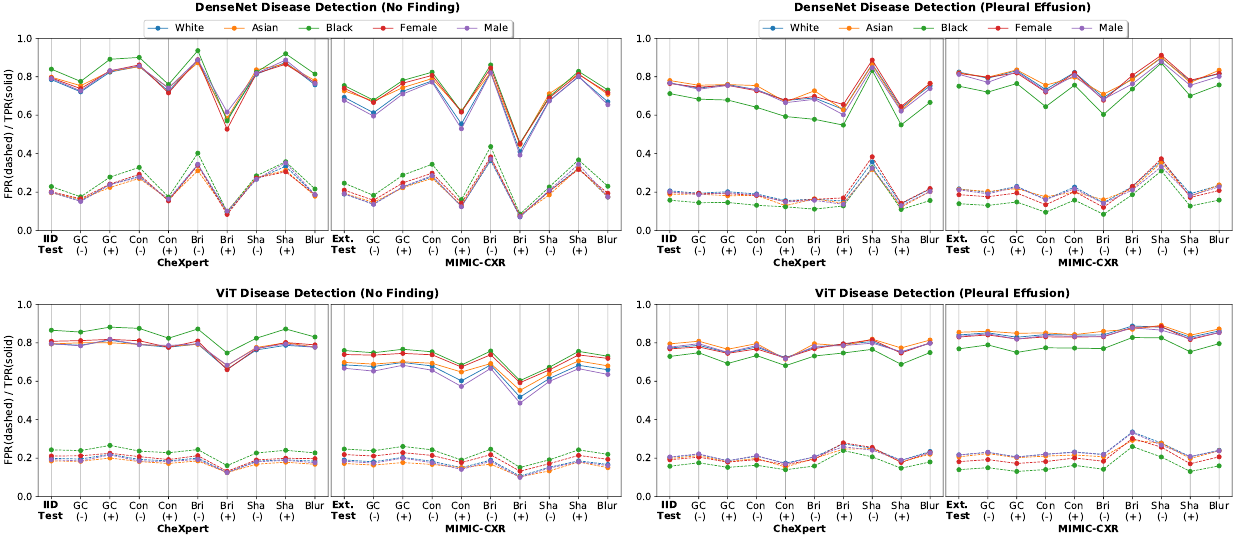}
\caption{TPR and FPR change across patient subgroups for a DenseNet (top) and ViT (bottom) disease detection model for a variety of stress tests on the development data (CheXpert) and the external validation data (MIMIC-CXR). Performance for `no-finding' on the left, and `pleural effusion' on the right.}
\centering
\label{fig:robustness_perturbation}
\end{figure*}

\begin{figure*}[!h]
\centering
\includegraphics[width=1\textwidth]{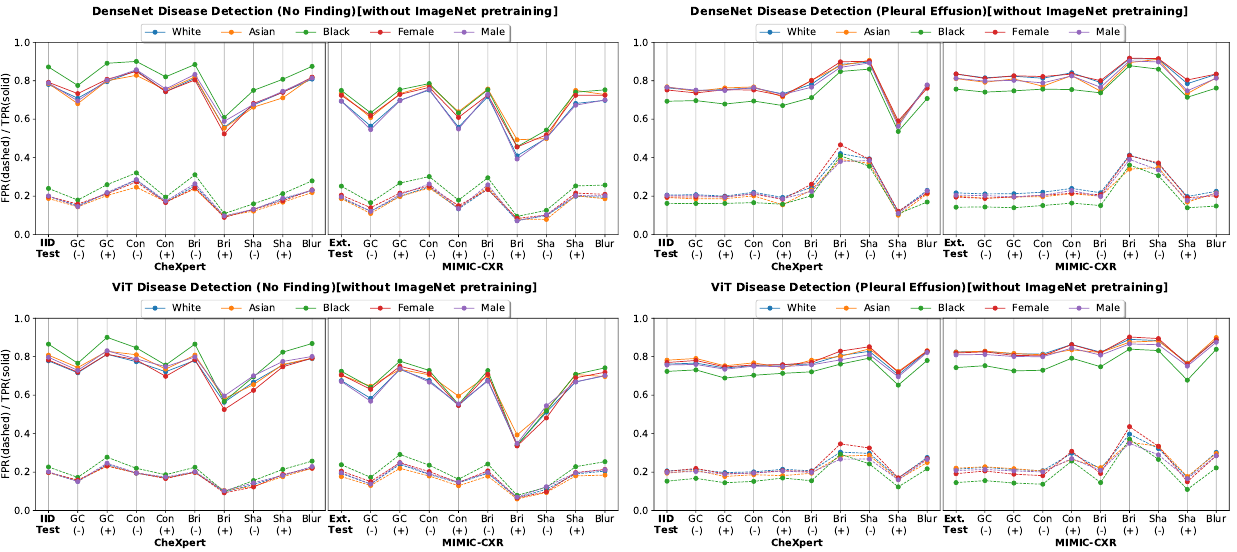}
\caption{TPR and FPR change across patient subgroups for a DenseNet (top) and ViT (bottom) disease detection model trained from scratch (without ImageNet pretraining).}
\centering
\label{fig:robustness_tpr_fpr_thr_wo_pretrain}
\end{figure*}

\begin{figure*}[!h]
\centering
\includegraphics[width=1\textwidth]{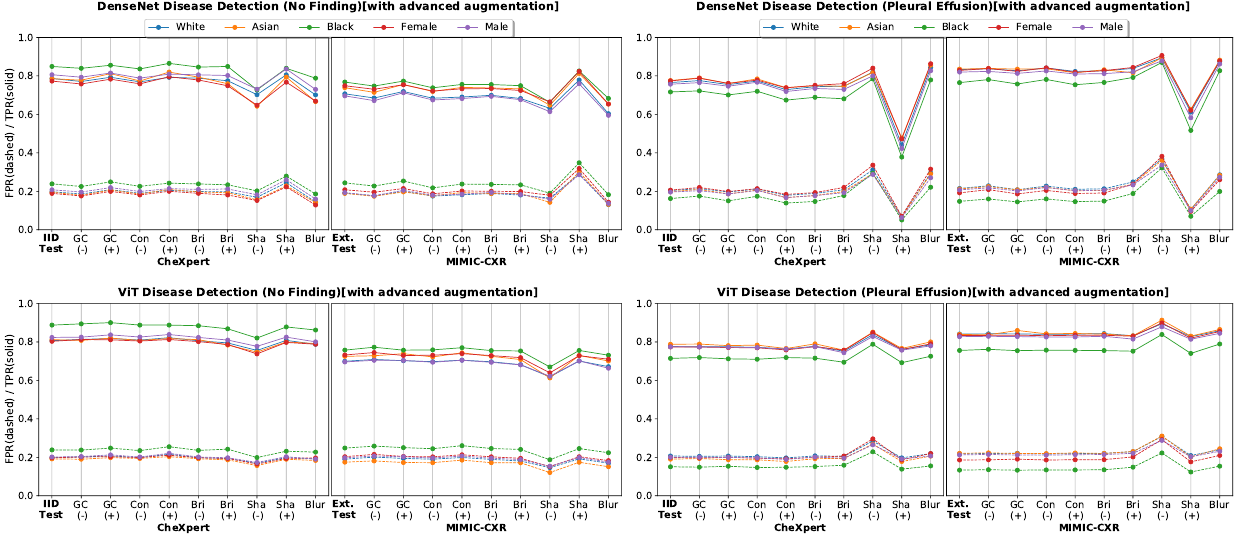}
\caption{TPR and FPR change across subgroups for DenseNet and ViT trained with advanced augmentations including gamma correction, contrast, and brightness.}
\centering
\label{fig:robustness_tpr_fpr_thr_aug}
\end{figure*}

\begin{figure*}[!t]
\centering
\includegraphics[width=1\textwidth]{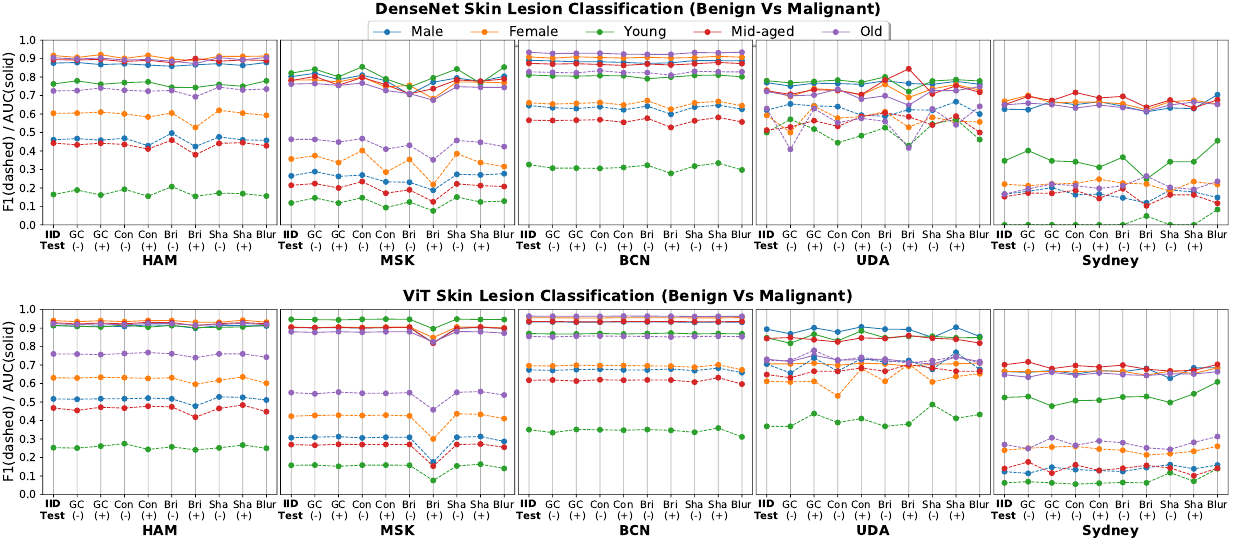}
\caption{F1 and AUC change across patient subgroups over five different test domains for a DenseNet (top) and ViT (bottom) disease detection model.}
\centering
\label{fig:robustness_auc_skin}
\end{figure*}

\begin{figure*}[!h]
\centering
\includegraphics[width=1\textwidth]{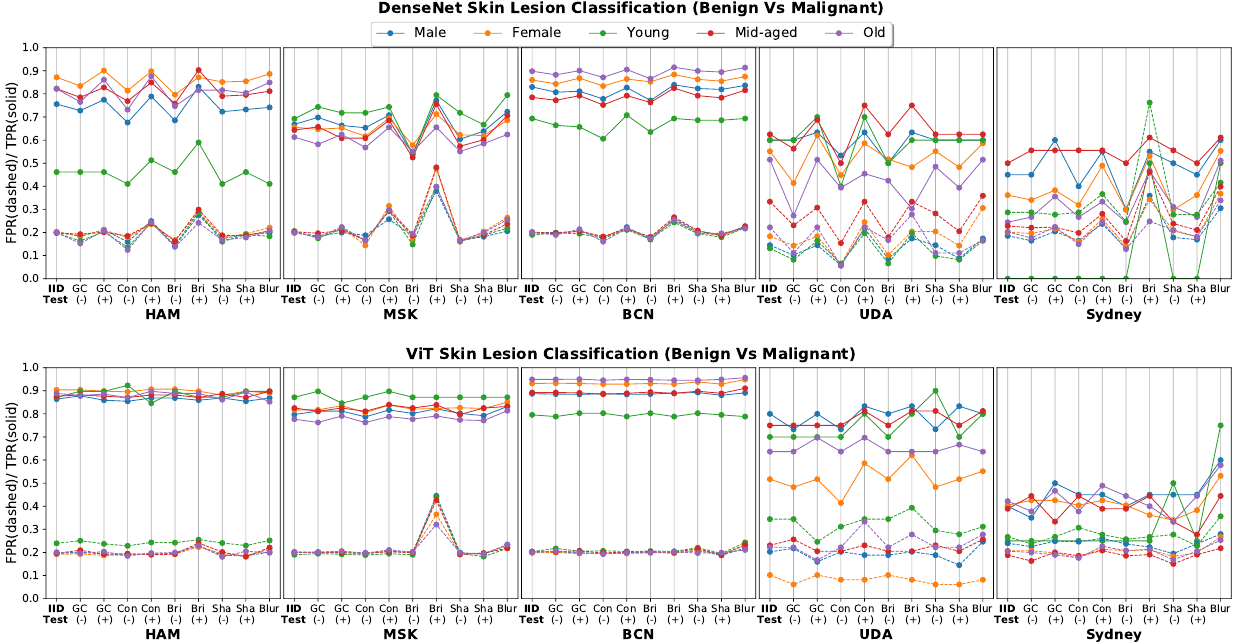}
\caption{TPR and FPR change across patient subgroups over five different test domains for a DenseNet (top) and ViT (bottom) skin lesion classification model.}
\centering
\label{fig:robustness_perturbation_skin}
\end{figure*}

The performance for the chest X-ray disease detection model on iid (CheXpert), external (MIMIC-CXR) test sets (without any perturbations applied) and the skin lesion classification on multi-site (ISIC) test sets are given in Table~\ref{tab:chest_xray}. To analyse and compare the robustness of the different models, we employ progressive stress testing via image perturbations. We apply 30 different image perturbations to assess model robustness. In the following we only choose one positive ($+2$) and negative ($-2$) adjustment for each bidirectional perturbation and only one positive ($+2$) adjustment for the unidirectional one (as in Fig~\ref{fig:perturbation_techs}), as we found that the change in performance is monotonic. The results of the stress testing of the chest X-ray disease detection models are shown in Fig~\ref{fig:robustness_auc} and Fig~\ref{fig:robustness_perturbation}. To investigate the somewhat surprising robustness of the ViT, we also trained both models from randomly initialized weights instead of using ImageNet pretraining, keeping all other settings unchanged as shown in Fig~\ref{fig:robustness_tpr_fpr_thr_wo_pretrain}. To further analyse the mixing effects of pretraining and data augmentation on model robustness, we utilized additional appearance augmentation techniques during training including gamma correction, contrast, and brightness (cf. Fig~\ref{fig:robustness_tpr_fpr_thr_aug}). We also observe a much higher degree of robustness of ViT compared to DenseNet across the different stress tests on skin lesion classification task as shown in Fig~\ref{fig:robustness_auc_skin} and Fig~\ref{fig:robustness_perturbation_skin}. Additional figures and observations on advanced augmentation, ResNet vs Swin-Transformer, and robustness in model calibration can be found in the supplementary materials.

From the experimental results, figures, and tables, our observation can be summarized as: (i) Transformer-based networks (ViT or Swin-Transformer) yield much more stable TPR/FPR performance across perturbations compared to the CNN-based networks (DenseNet, ResNet) (as shown in Fig~\ref{fig:robustness_perturbation}). (ii) The effect of the perturbations is larger for certain classes, which highlights the importance of using class-specific stress testing for complete robustness analysis (as shown in Fig~\ref{fig:robustness_perturbation}). (iii) In terms of the AUC metric alone, all types of network architectures show a robust performance across stress tests in most of the cases but varying TPR/FPR performance under the same threshold (as shown in Fig~\ref{fig:robustness_auc} and TABLE~\ref{tab:chest_xray}). (iv) ViT substantially degrades in robustness if training without ImageNet pretraining weights, which indicates that the ImageNet pretraining is not only important to obtain downstream robustness but that the exact nature of pretraining might be important (as shown in Fig~\ref{fig:robustness_tpr_fpr_thr_wo_pretrain}). (v) Advanced augmentations improve the robustness, in particular for the DenseNet in corresponding perturbations, but ViT appears overall more robust even to perturbations outside the data augmentation strategy (as shown in Fig~\ref{fig:robustness_tpr_fpr_thr_aug}). (vi) In terms of model robustness under domain shift, iid and external test sets CheXpert and MIMIC-CXR seem largely consistent with similar fluctuations (as shown in Fig~\ref{fig:robustness_perturbation}); however, we observe significant changes in overall performance across the test domains (HAM, MSK, BCN, UDA, and SYD) for skin lesion classification (as shown in Fig~\ref{fig:robustness_perturbation_skin}). (vii) The progressive stress testing also can highlight the subgroup disparity for underrepresented groups (e.g., Black patients) (as shown in Fig~\ref{fig:robustness_perturbation}) and negatively affect the group of female patients for some architectures with advanced data augmentation (as shown in Fig~\ref{fig:robustness_tpr_fpr_thr_aug}).

\section{Discussion and conclusion}

In this work, we have explored progressive stress testing as a comprehensive tool to assess robustness in image classification models. Our work extends previous frameworks, arguing that it is important to include subgroup analysis and class-specific performance monitoring. The example applications of chest X-ray disease detection and skin lesion classification highlight the value of progressive stress testing to reveal robustness characteristics that otherwise remain hidden when using traditional test set evaluation. We found differences in robustness between state-of-the-art neural network architectures. We found a connection between ImageNet pretraining and downstream robustness which potentially has a larger contribution than the choice of the neural network architecture. The transformer-based models, ViT and Swin-Transformer, appear generally more robust than the CNN-based models, DenseNet and ResNet. The future direction of this work is to consider adversarial attacks to design adversarial stress testing.

\section*{Acknowledgements}
This project has received funding from the European Research Council (ERC) under the European Union's Horizon 2020 research and innovation programme (grant agreement No 757173, project MIRA, ERC-2017-STG).

\bibliographystyle{splncs04}
\bibliography{mybibliography}
\clearpage
\appendix



%
%
%

%

\begin{center}
 \textbf{\large Supplementary Materials}   
\end{center}

\begin{figure*}[!h]
\centering
\includegraphics[width=1\textwidth]{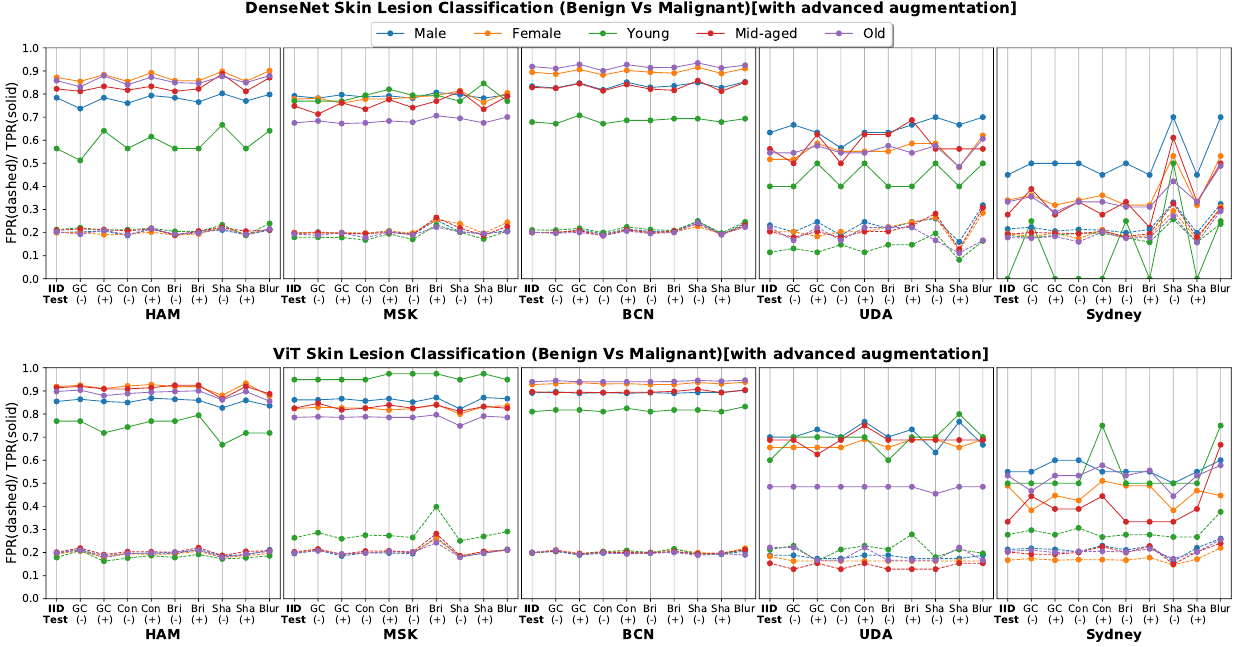}
\caption{Skin lesion classification trained with advanced augmentations including gamma correction, contrast, and brightness. We observe that the use of data augmentation can hurt robustness on perturbations in some domains, which reveals the importance of using carefully designed data augmentation strategies.}
\centering
\label{fig:robustness_aug_skin}
\end{figure*}

\begin{figure}[!h]
    \begin{minipage}{0.49\linewidth}
        \includegraphics[width=1\textwidth]{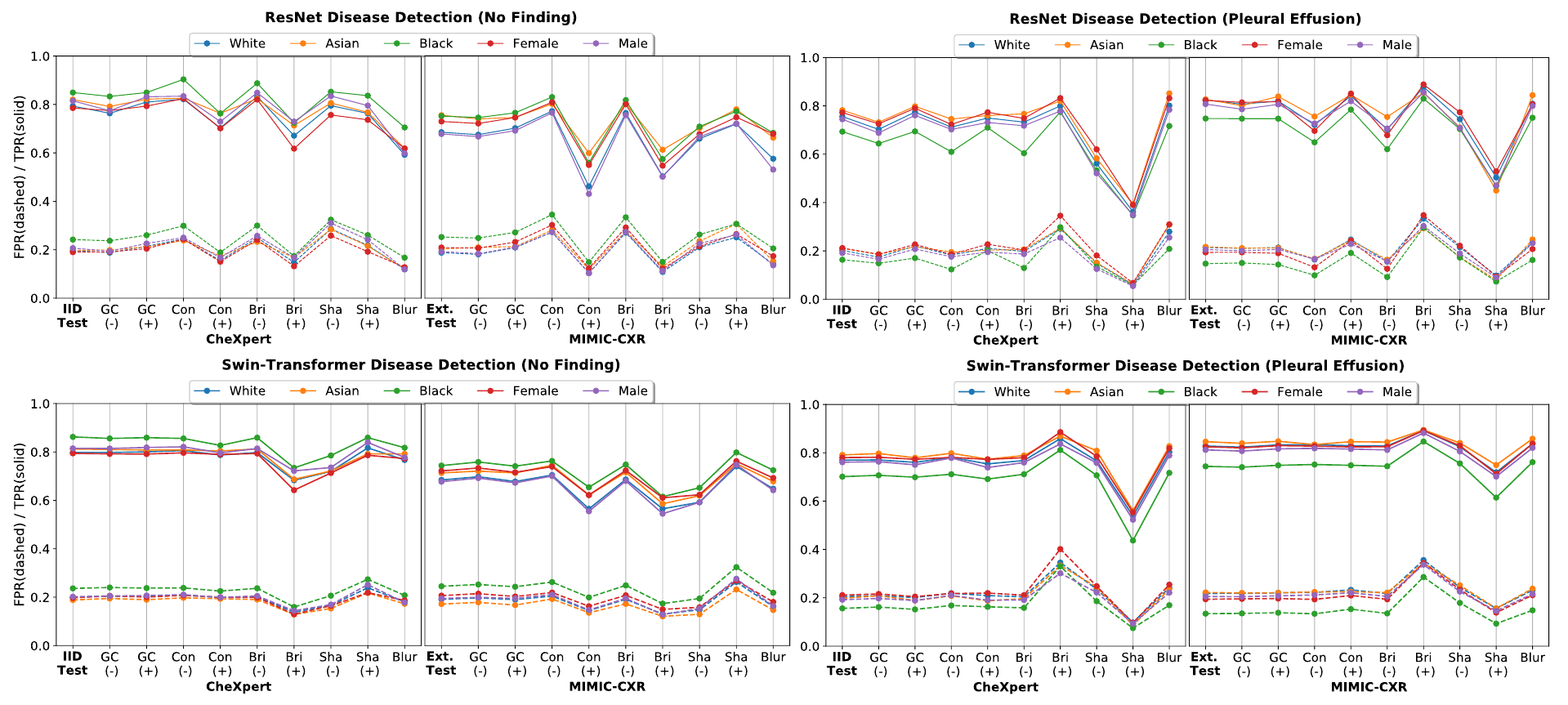}
        \caption{TPR and FPR change across patient subgroups for a ResNet (top) and Swin-Transformer (bottom). Similar to ViT, Swin-transformer also yields much more stable TPR/FPR performance across perturbations compared to the ResNet.}
    \end{minipage}\hfill
    \begin{minipage}{0.49\linewidth}
        \includegraphics[width=1\textwidth]{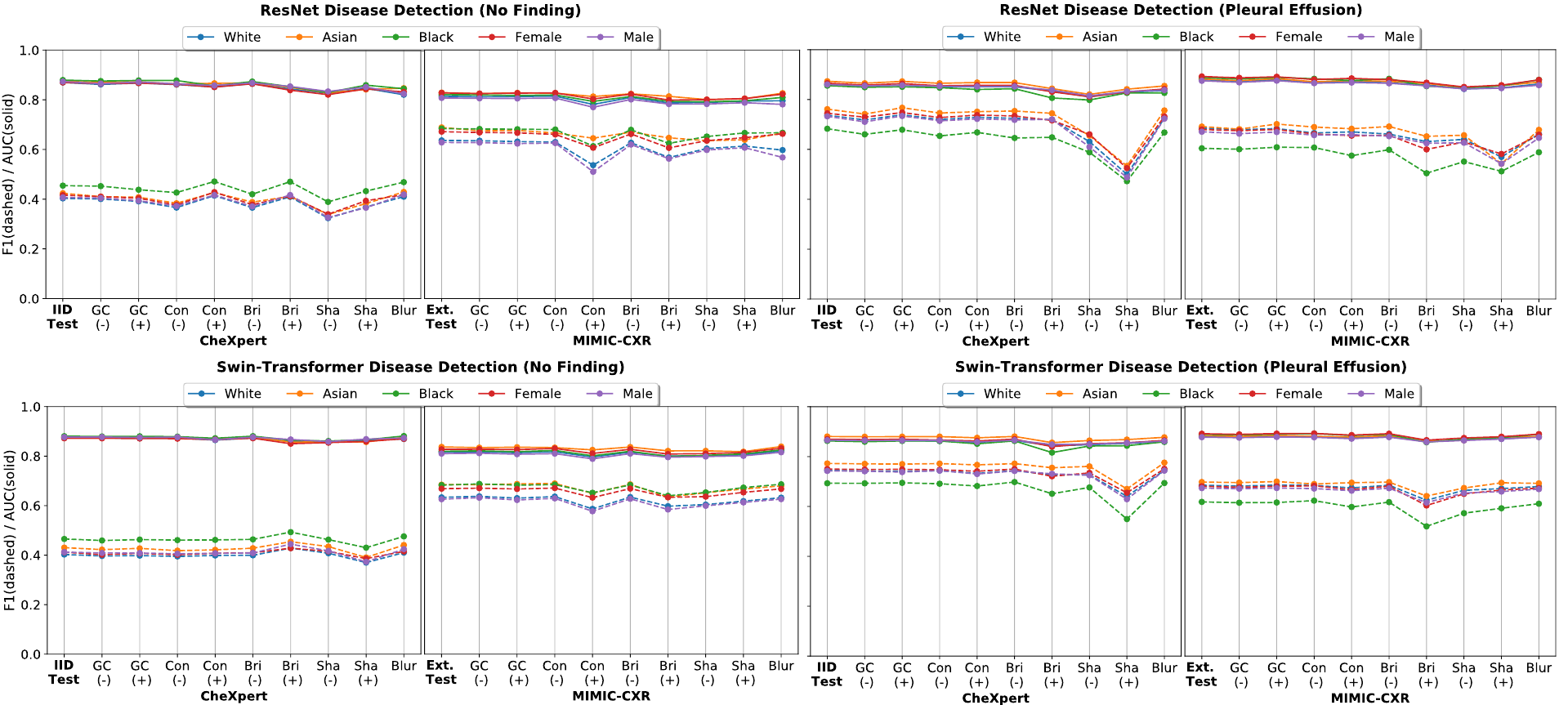}
        \caption{{F1 and AUC change for `no-finding' and  `pleural effusion' across patient subgroups for a ResNet (top) and Swin-Transformer (bottom) disease detection model. The metrics do not reveal much about robustness.}}
    \end{minipage}
\end{figure}

\begin{figure}[!t]
    \begin{minipage}{0.49\linewidth}
        \includegraphics[width=1\textwidth]{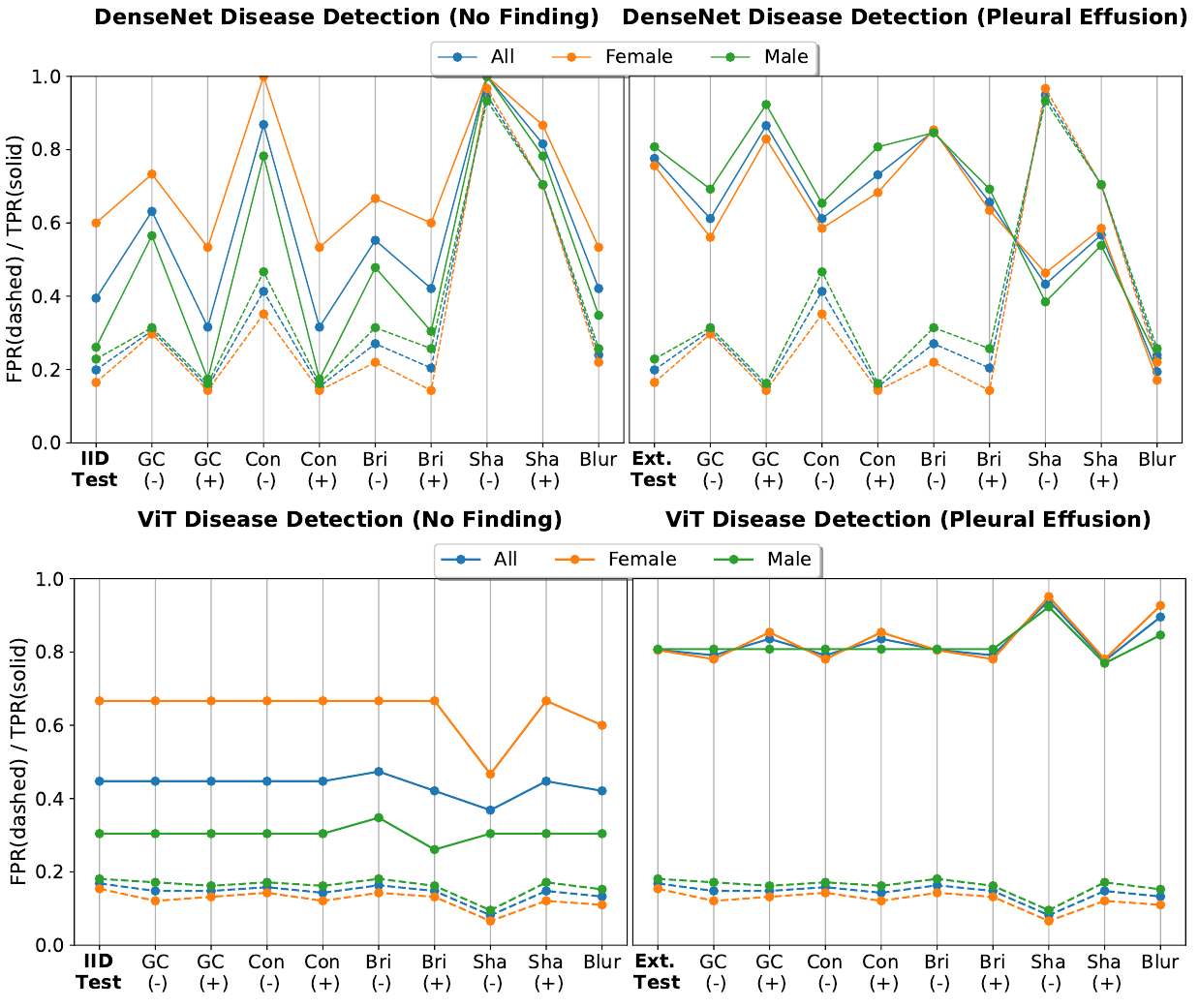}
        \caption{{TPR and FPR change for `no finding' and `pleural effusion' across patient subgroups for a DenseNet (top) and ViT (bottom) disease detection model for CheXphoto dataset. Similar to earlier findings, transformer-based architecture is showing better robustness over CNN.}}
    \end{minipage}\hfill
    \begin{minipage}{0.49\linewidth}
        \includegraphics[width=1\textwidth]{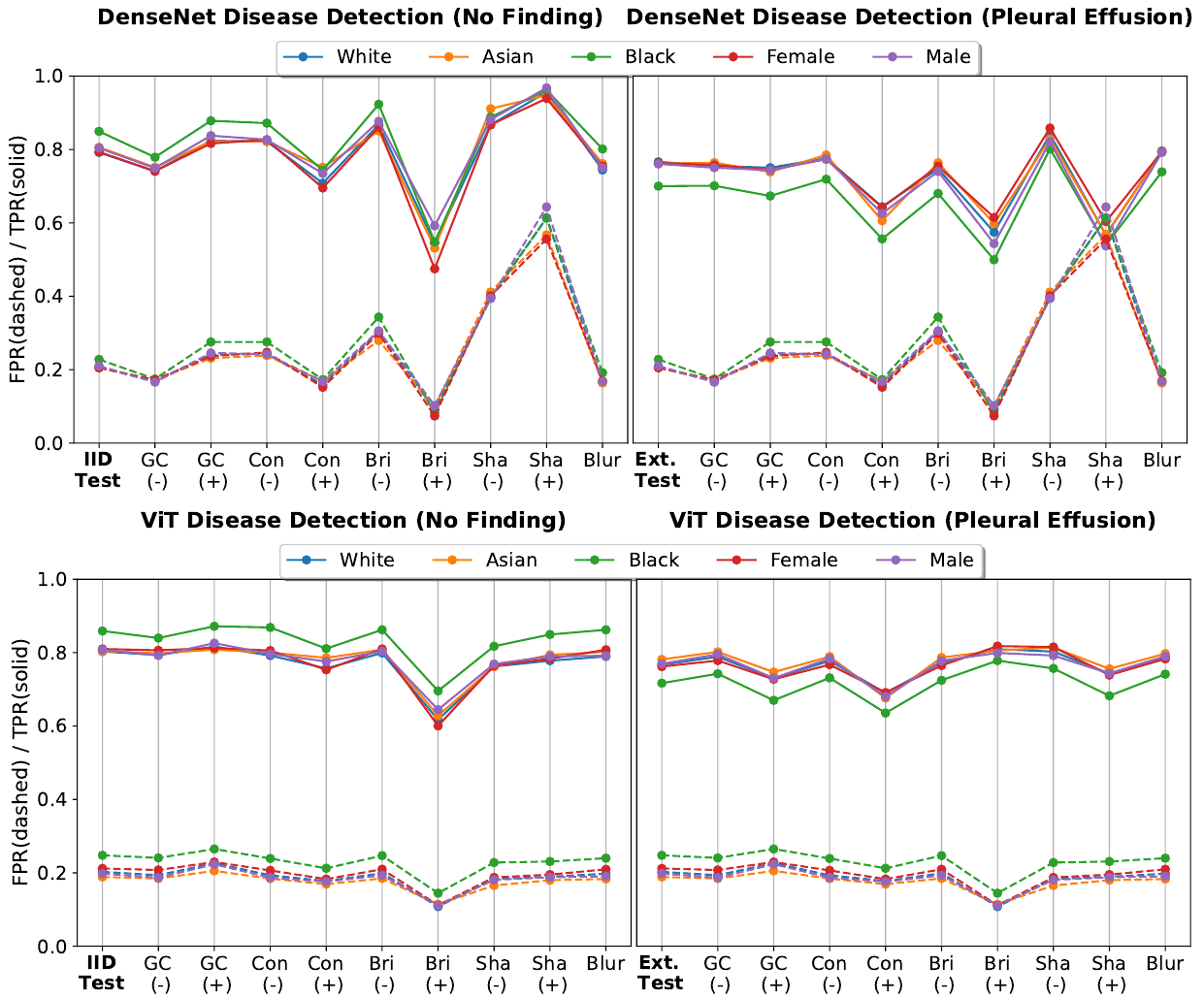}
        \caption{TPR and FPR change for `no finding' and `pleural effusion' across patient subgroups for a DenseNet (top) and ViT (bottom) disease detection model for CheXpert dataset with different random seeds. The overall trend is still similar compare to the main manuscript Fig 3.}
    \end{minipage}
\end{figure}

\begin{figure}[!h]
\includegraphics[width=1\textwidth]{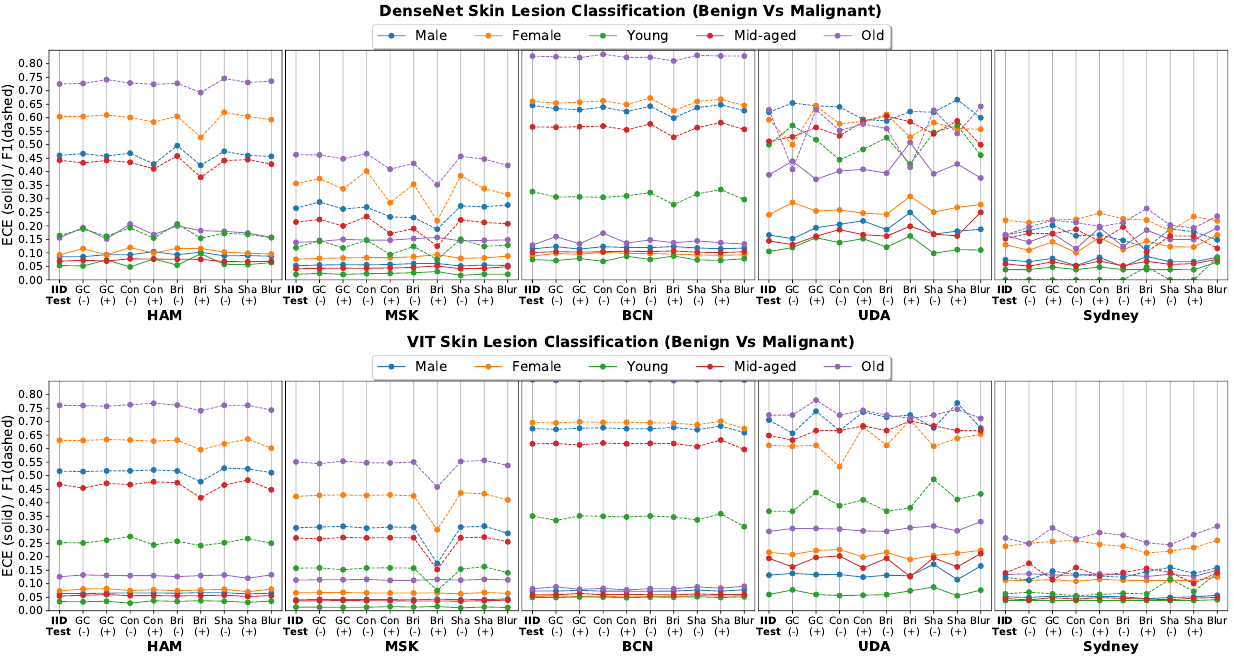}
\caption{Expected calibration error (ECE) and F1 change across patient subgroups over five different test domains for skin lesion. The changes of ECE, as well as F1 in DenseNet, are higher than ViT. In fact, there are changes in ECE (increasing mostly) for HAM and BCN where the F1 score is almost constant under the perturbation of GC(+), Con(-), and Bri (-) for the prediction in DenseNet. Interestingly, the subgroup of Old age patients showed higher calibration errors for both architectures, which obtained higher performance in the F1 score.}
\centering
\label{fig:f1_ece_skin}
\end{figure}


\end{document}